\documentclass[aip,amsmath,amssymb,reprint]{revtex4-2}
\usepackage{xcolor}
\usepackage{multirow}
\usepackage{times}
\usepackage{graphicx}
\usepackage{dcolumn}
\usepackage{bm}
\usepackage{hyperref}

\begin{document}

\preprint{APS/123-QED}

\title{Millikelvin magnetometry up to 30~T using a multi-cross micro-Hall sensor}

\author{Bingbing Tong} 
\altaffiliation{These authors contributed equally to this work.}
\affiliation{Beijing National Laboratory for Condensed Matter Physics and Institute of Physics, Chinese Academy of Sciences, Beijing 100190, China}

\author{Yinina Ma} 
\altaffiliation{These authors contributed equally to this work.}
\affiliation{Beijing National Laboratory for Condensed Matter Physics and Institute of Physics, Chinese Academy of Sciences, Beijing 100190, China}

\author{Zhenhua Zhao} 
\affiliation{School of Physical Sciences, University of Chinese Academy of Sciences, Beijing 100190, China}

\author{Shuyu Zheng} 
\affiliation{Beijing National Laboratory for Condensed Matter Physics and Institute of Physics, Chinese Academy of Sciences, Beijing 100190, China}

\author{Li Lu} 
\affiliation{Beijing National Laboratory for Condensed Matter Physics and Institute of Physics, Chinese Academy of Sciences, Beijing 100190, China}
\affiliation{School of Physical Sciences, University of Chinese Academy of Sciences, Beijing 100190, China}
\affiliation{Hefei National Laboratory, Hefei, Anhui 230088, China}

\author{Peijie Sun} 
\email{pjsun@iphy.ac.cn}
\affiliation{Beijing National Laboratory for Condensed Matter Physics and Institute of Physics, Chinese Academy of Sciences, Beijing 100190, China}
\affiliation{School of Physical Sciences, University of Chinese Academy of Sciences, Beijing 100190, China}

\author{Gang Li} 
\email{gli@iphy.ac.cn}
\affiliation{Beijing National Laboratory for Condensed Matter Physics and Institute of Physics, Chinese Academy of Sciences, Beijing 100190, China}
\affiliation{School of Physical Sciences, University of Chinese Academy of Sciences, Beijing 100190, China}


\begin{abstract}
We report a multi-cross micro-Hall sensor fabricated from an MBE-grown GaAs/AlGaAs heterostructure for high-sensitivity magnetometry. The device enables dc magnetization measurements at temperatures down to $\sim$15~mK and in dc magnetic fields up to 30 T. Bulk samples with dimensions ranging from 100~$\mu$m to 1~mm can be measured with a resolution better than 1$~\mu$T with a 1~$\mu$A excitation at 13.77 Hz. Test-of-performance measurements were carried out by probing the field-dependent magnetization plateaus of the triangular-lattice quantum magnet K$_{2}$Co$_{2}$(SeO$_{3}$)$_{3}$ at 20~mK, the antiferromagnetic transition of K${_3}$GdTe${_2}$O${_9}$ near 220 mK via temperature-dependent magnetization, and the quantum oscillations in a ruthenium single crystal. A simple and reliable sample-mounting setup that ensures reproducible magnetic measurements is provided.

\end{abstract}


\maketitle


 \section{Introduction}
Magnetic field sensing and magnetic characterization of solid-state materials are central to both fundamental and applied physics, from condensed matter research to quantum technologies. For the measurement of bulk or macroscopic magnetic response, established techniques detecting magnetic flux or its variation, such as superconducting quantum interference device (SQUID) magnetometry and vibrating-sample magnetometry (VSM)\cite{VSM-review}, are commercially available , typically enabling measurements down to 1.8 K at fields up to $\sim$ 16 T provided by superconducting magnets. Extending the accessible parameter space to lower temperatures and higher magnetic fields has been a long-standing goal in frontier research. For example, commercial Magnetic Property Measurement System (MPMS, Quantum Design) equipped with a stand-alone iQuantum $^3$He insert (iHelium3) can reach temperatures as low as 0.4 K (ref.~\cite{QD_iHelium3_2015}), and VSM operating up to $35$ T (ref.~\cite{NHMFL_VSM}) is available in a few national labs equipped with water-cooled resistive magnets. Alternatively, the Faraday force method, particularly when implemented with a capacitive readout, has proven to be a powerful approach for high-sensitivity magnetic measurements at millikelvin temperatures and high magnetic fields, as demonstrated by Sakakibara et al.\cite{Sakakibara1994}. Recent interest in emergent quantum magnetism, unconventional superconductivity, and topological states in various quantum materials has raised an urgent demand for convenient magnetic measurements under extreme conditions \cite{Zhou2017, Grigorenko2001, Khatua2023, NBCP-Nature, KCSO-arxiv}, yet such measurements remain difficult to conduct or have highly limited accessibility.

\begin{figure*}[t]
\includegraphics[width=0.85\textwidth]{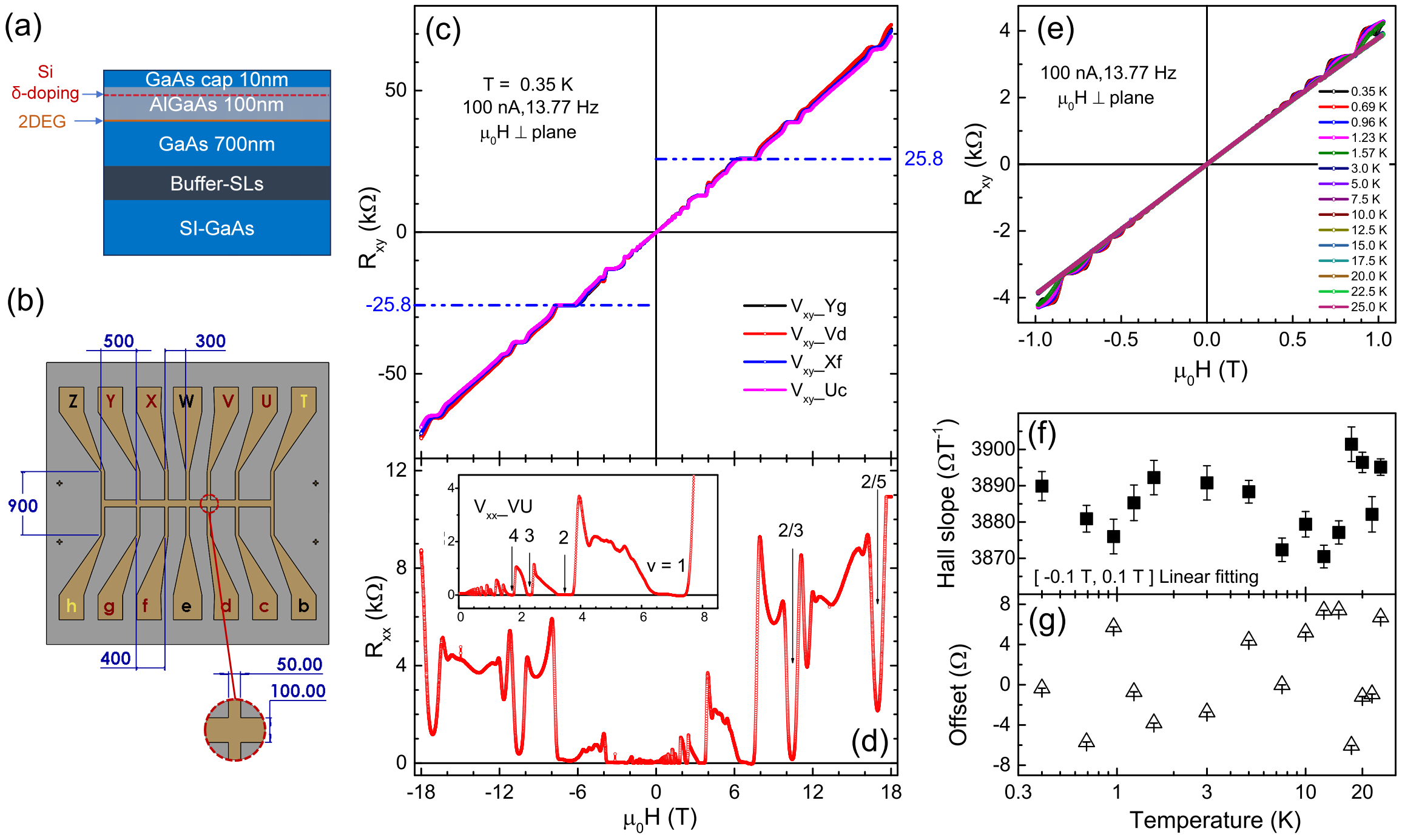}
\caption{\label{EmptyDevice}
 Characterization of a multi-cross micro-Hall sensor. (a) shows the schematic drawing of the GaAs/AlGaAs heterostructure, not in actual ratio. On top of the substrate, the epitaxial stack comprises a buffer layer consisting of $300$ nm GaAs and a 50-period superlattice, followed by a $700$ nm undoped GaAs layer, a $100$ nm AlGaAs layer, and a $10$ nm GaAs protective cap. In this structure, the spacer thickness between the Si $\delta$-doping layer and the underlying 2DEG is $80$ nm. (b) gives the layout and dimensions of the Hall device with contacts assignments. The enlarged view shows the active Hall area. Units in $\mu$m. (c) shows the transverse response of four pairs of Hall contacts at $0.35$ K and in external magnetic fields up to $18$ T, and (d) is the longitudinal response from one pair of electrodes with the zoomed-in low-field part as an inset. Integer and fractional quantum Hall features are marked. The isothermal responses in a perpendicular field mainly within the low field linear region are given in (e), the corresponding slopes and offset (intercepts at zero field) by linear fitting are in (f) and (g), respectively.}
\end{figure*}

Complementing these aforementioned flux- and force-based techniques, Hall magnetometers \cite{ Konczykowski1991, qd2008, Karsenty2020,Novoselov2003Submicron,lefebvre2009local,james1997field} can provide a conceptually simple and versatile route to probe dc magnetization by detecting the local magnetic field at the sensor. When a sample is magnetized in an external magnetic field, it generates a stray magnetic field outside its volume. A Hall sensor placed nearby detects the component of this stray field, $B_{\rm stray}$, perpendicular to its sensing plane. For a given, fixed sample geometry, $B_{\rm stray}$ is directly proportional to the sample's total magnetic moment, and hence to its magnetization $M$: $B_{\rm stray}$ $\propto$ $M$. If the Hall response is linear in the local magnetic field, the measured Hall resistance becomes proportional to the sample's magnetization, $R_{\rm xy}$ (= $V_{\rm xy}$/$I$) $\propto$ $B_{\rm stray}$ $\propto$ $M$.

With advances in two-dimensional electron gas (2DEG) device fabrication techniques, there is considerable flexibility in designing a Hall sensor in a compact form. Furthermore, a Hall sensor based magnetometer has no moving parts, thus can be easily integrated into cryogenic fridges with limited space, including challenging environments of very low temperatures as well as high magnetic fields\cite{CeRuAs122, NBCP-PNAS}. However, realizing a practical Hall magnetometer is a substantial undertaking. The design and growth of a 2DEG system with desired mobility and charge carrier density, the design, patterning, and fabrication of micrometer-scale Hall crosses, as well as the scheme to couple a magnetized sample to the Hall sensor and the procedure to mount samples for reproducible measurements, are mutually interconnected tasks. For these reasons, high-sensitivity micro-Hall sensors specially designed for solid-state magnetometry at extreme conditions remain rare.\\

In this work, we report the characterization of a micro-sized multi-cross Hall sensor based on 2DEG confined in a GaAs/AlGaAs heterostructure, all fabricated in-house. A stacking method is utilized to mount samples firmly in desired orientation and in close proximity to the Hall sensor. A series of temperature and magnetic field dependent magnetization measurements were carried out at the Synergetic Extreme Condition User Facility (SECUF), with temperature down to $\sim$ 15~mK, and magnetic field up to $30$ T. The robust and consistent performance of the Hall magnetometer is demonstrated by data on K$_{2}$Co$_{2}$(SeO$_{3}$)$_{3}$, K${_3}$GdTe${_2}$O${_9}$, and element Ruthenium (Ru). These results establish the effectiveness and sensitivity of the device and position it as a practical tool ready for broader use, with directions for further performance improvements.

\section{Devices and Experiments}

\noindent \textbf{Fabrication and characterization of the Hall sensor.} The GaAs/AlGaAs heterostructure was grown by molecular beam epitaxy (MBE) on an undoped GaAs substrate with the structure shown in Fig.~\ref{EmptyDevice} (a)\cite{1529831092702-21246461}. Chips of 4 $\times$ 4 mm$^{2}$ were cut from the grown wafer and then patterned into a Hall bar by photolithography and wet etching, each including an array of five Hall probes (labeled Yg, Xf, We, Vd, and Uc), as visualized in Fig.~\ref{EmptyDevice}(b). The active Hall probe area of 100 $\times$ 50 $\mu$m$^2$ and Hall probe spacing were determined by taking into account the physical characteristics of the 2DEG from prior knowledge, as well as typical dimension ($\sim$ 300 $\mu$m) of samples of interest, to avoid the nonlinear behavior of Hall resistance with magnetic field due to ballistic transport effects \cite{jouault2009ballistic} and non-uniform magnetic field\cite{gallagher2001electrical}. Ohmic contacts were formed using In-Sn alloy annealed at 450 $^\circ\mathrm{C}$ in N$_{2}$/H$_{2}$ forming gas. \\

\begin{figure*}
\includegraphics[width=0.85\textwidth]{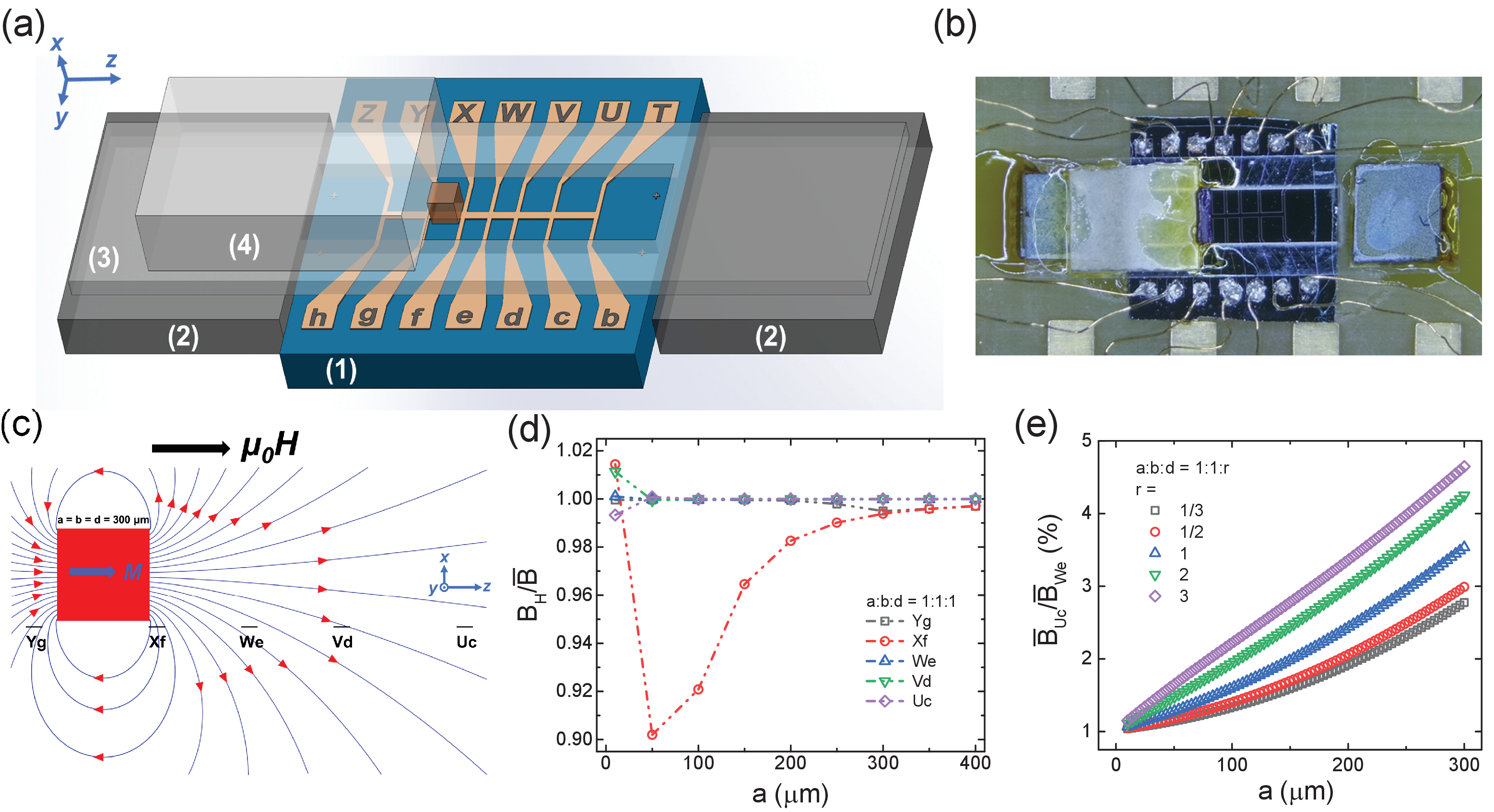}
\caption{\label{Device}
The micro-Hall magnetometer device setup and Hall response simulation. (a) is a schematic cartoon of device assembly, in which (1) is the multi-cross micro-Hall chip (4$\times$4 mm$^{2}$). (3) is a slotted sapphire chip that is glued to the spacers (2) to form a flat surface, and (4) is second sapphire chip with well-defined, normal surfaces for mounting samples. Labeling of Hall contacts is carried over from Fig.~\ref{EmptyDevice}. (b) is a photograph of the actual device mounted on a 16-pin, dual in-line socket. Simulations were carried out by assigning the external magnetic field along the z-axis of the coordinate system, and the sample was represented by a cube with dimensions a, b, and d along x, y, and z-axis, respectively. The sample is placed 20 $\mu$m above the Hall chip. Homogeneous magnetization was assumed. (c) illustrates the distribution of magnetic flux from a magnetized sample with respect to Hall probes. (d) gives the ratio of the magnetic flux calculated from Hall voltage to the averaged magnetic flux experienced by a Hall cross as a function of sample size. (see text) (e) is the ratio of the averaged magnetic flux from two Hall crosses, plotted as a function of sample size.}
\end{figure*}

To benchmark the sensor, low-temperature transport measurements were performed in a $^3$He cryostat (0.35--25~K), with magnetic fields up to 18~T applied perpendicular to the 2DEG plane. Standard lock-in techniques were used with an AC excitation current of 0.1--1~$\mu$A at frequencies below 200~Hz. The Hall resistance $R_{\rm xy}$ of an empty Hall device measured at $T$ = 0.35~K is shown in Fig.~\ref{EmptyDevice}(c). The $R_{xy}(\mu_0H)$ data at four different Hall crosses were recorded simultaneously. The four curves show good agreement, the deviation of any individual curve from the average is within 5~\% in the whole field range. Integer quantum Hall plateaus and fractional quantum Hall states at $\nu=2/3$ ($B\simeq$ 10.4~T) and $\nu=2/5$ ($B\simeq$ 17.3~T) are clearly resolved, attesting to the high quality of the heterostructure. However, asymmetric longitudinal magnetoresistance $R_{xx}$ with respect to magnetic field still exists, as shown in Fig.~\ref{EmptyDevice}(d), possibly due to spatial inhomogeneity in the 2DEG in the absence of illumination treatment. The electron density and low temperature mobility were calculated from both $R_{xy}$ and $R_{xx}$. We obtain \(n_e = 1.67 \times 10^{11}\) cm\(^{-2}\) and \(\mu=2.45 \times 10^6\) cm\(^2\)V\(^{-1}\)s\(^{-1}\), respectively. The corresponding electron mean free path is about 16.6 $\mu$m.

A set of isothermal Hall resistance curves within $\pm$1 T from 0.35 to 25 K are displayed in Fig.~\ref{EmptyDevice}(e). The response is linear in field for $|B| \lesssim 0.2$~T at all measured temperatures, with the linear fitting slopes varying within 0.4\%, as shown in Fig.~\ref{EmptyDevice}(f). The extracted temperature-independent field responsivity is 3.885~$\Omega$/mT (388.5~nV/mT at 0.1~$\mu$A excitation). Besides, the absolute value of offset at zero field from unconstrained linear fitting, although mainly attributable to instrumental effects, is less than $8\ \Omega$ and shows a weak dependence on temperature (Fig.~\ref{EmptyDevice}(g)). These robust characteristics of the Hall sensor response form the basis for subsequent temperature-dependent magnetization measurements down to millikelvin range.  \\

\noindent \textbf{Sample mounting and alignment.}
Given the linear range of the fabricated 2DEG Hall device and our intention to utilize it for magnetization measurements down to millikelvin temperatures and/or up to high magnetic fields, a configuration with external field $H_{\rm applied}$ parallel to the 2DEG plane is employed. Fig.~\ref{Device}(a) shows a schematic of the Hall sensor assembly with a sample mounted. The key advantage of the configuration is that the sample can be placed in close proximity to the surface of the Hall device without direct contact. Two spacers, labeled as (2), made from scrap from the GaAs substrate, are used to support the slotted sample-mounting sapphire frame (3). By adding a layer of varnish-soaked cigarette paper to the spacers, a defined, $\sim$20 $\mu$m gap is formed between the top surface of (1) and the bottom surface of (3). The Hall chip and the spacers are fixed in place. To fully assemble a device, the sample is first glued in the desired orientation to the positioning sapphire chip (4) with well-defined, normal surfaces, then the relative position of (4) to (3) and (3) to (1) could all be finely adjusted to position the sample relative to the Hall sensing elements, which serve as positioning references. Finally, (3) and (4) are anchored with varnish. This procedure is simple yet flexible. As the sample and the Hall device are kept separate, the risk of scratching or straining the surface of Hall chip is reduced. Fig.~\ref{Device}(b) is a photograph of the assembly. The overall footprint is about 12$\times$6~mm$^{2}$, which can easily fit into the standard 16-pin DIP sockets used in our $^{3}$He and $^{4}$He fridges, as well as into a custom 24-pin square socket for our top-loading $^3$He-$^4$He dilution refrigerator for measurements in millikelvin range. For a quick turnaround in experiments, the sample could be taken off together with the sapphire frame, without disturbing the Hall chip and all electrical contacts.

The Hall response of the parallel external field setup was simulated. Following the coordinates defined in Fig.~\ref{Device}(a), the external magnetic field is along z-axis, while the Hall device lies in yz plane. The Hall device responses to the x-axis component of the stray field of the magnetized sample, $B_{\rm{stray}-x} (y,z)$, which has a spatial distribution in the yz plane. For simplicity, homogeneous magnetization is assumed while the demagnetization factor is neglected. The spatial distribution of magnetic flux can be calculated analytically\cite{Field-Distri}. Figure~\ref{Device}(c) depicts the distribution of magnetic flux from a 300~$\mu$m cubic sample viewed along y-axis. It is clear that even within the active Hall sensing area, $B_{\rm{stray}-x} (y,z)$ varies and its spatial average, denoted $\bar{B}$, is the relevant quantity of interest. Meanwhile, $B_{H}$ is deduced from the measurable quantity, Hall voltage $V_{H}$, which is calculated following the classical Drude model\cite{Hall-Calc} as the difference between the average electric potential along the opposing voltage contacts of the Hall bar cross. Assumptions used include that the diagonal components of the resistivity tensor are constant, while the off-diagonal components, the Hall resistance $R_{H}(y,z)$, are proportional to $B_{\rm{stray}-x} (y,z)$. The carrier concentration and mobility are taken with actual values for the device, and an effective, saturating magnetization corresponding to 0.375 T is used. Figure~\ref{Device}(d) shows the ratio of $B_{H}/\bar{B}$ as a function of sample size. Except for the Hall sensing element closest to the sample, at the other positions the magnetic flux intensity deduced from Hall voltage is almost identical to the averaged magnetic flux intensity, which justifies the design of the Hall device. The ratio of the averaged magnetic flux intensity from the farthest Hall element to that from the third farthest one, is shown in Fig.~\ref{Device}(e). Even by increasing the sample dimension along the external field direction to 900 $\mu$m, the averaged magnetic flux intensity at the farthest Hall element remains less than 5 percent of that at 1000 $\mu$m away . It is expected and forms the base that the Hall signal from the Hall cross farthest away from the sample can be used as background.

Moreover, some physical characteristics of 2DEG, such as Fermi contour\cite{ohtsuka1998distortion}, electron scattering, spin polarization\cite{dolgopolov2000magnetoresistance}, and the $g$-factor\cite{liang2001spin}, are weakly affected by a parallel field,  background subtraction can eliminate the contributions from these effects to the Hall response of the sample, as well as from the systematic drifting of a cryogenic system.
Besides, the background Hall channel is used as a measure to align the device in parallel with the external magnetic field. For the device loaded into a rotator probe, the parallel position can be roughly reached by finding the Hall voltage zero at a constant external field. Thanks to the large Hall response and tiny offset of the device, this method could already reach a misalignment of less than 1 degree, while further alignment could be done by a series of field sweeps in the low field linear range and finding the linear slope minimum of Hall signal.

\section{Results and Discussion}

\begin{figure*}
\includegraphics[width=0.85\textwidth]{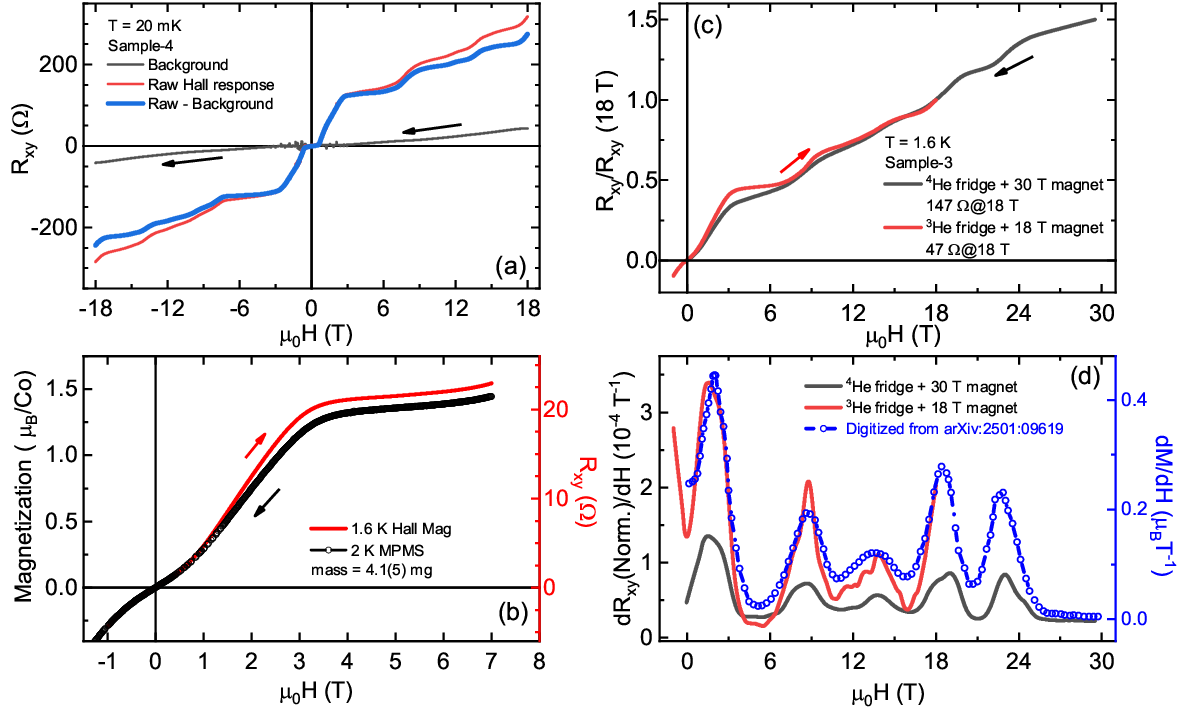}
\caption{\label{KCSO}
Isothermal magnetization of K$_2$Co$_2$(SeO$_3$)$_3$ with the external magnetic field along the crystalline c-axis. Arrows in panels indicate the field sweeping direction. (a) Measurement by the Hall magnetometer in a top-loading dilution refrigerator at base temperature of $20$ mK. Raw data and background-subtracted data are all presented. (b) Comparison of the Hall resistance $R_{\rm xy}$ at $1.6$ K with the direct dc magnetization $M$ measured by MPMS at similar temperature. It is noted that the two measurements were on different crystals, and the piece used in MPMS measurement was larger. (c) Comparison of normalized Hall resistance $R_{xy}(B)$ of one sample/device combination measured in two different low-temperature, high-magnetic field systems. (d) Corresponding differentiation of the data in (c), and high field data digitized from Ref.\cite{KCSO-arxiv} are included for comparison.
}
\end{figure*}

\noindent \textbf{Field-dependent magnetization.} One important envisioned application of the fabricated micro-Hall device is probing frustrated quantum magnetism at millikelvin temperatures and high magnetic fields. K$_2$Co$_2$(SeO$_3$)$_3$ (KCSO), a spin-1/2 dimer compound with a bilayer triangular-lattice structure, is one such material that has attracted widespread interest recently\cite{KCoSeO-Cava, KCSO-arxiv}. In this compound, the Co$^{2+}$ (3d$^7$) ions carry an effective spin-1/2 at low temperatures due to strong spin-orbit coupling and the low-symmetry crystal electric fields. The magnetization displays a sequence of plateaus at fractional values with increasing field before reaching saturation. There is no consensus on the magnetic field-temperature phase diagram yet, while a potential Berezinskii-Kosterlitz-Thouless (BKT) phase extending to well below $1$ K at finite fields has been suggested.\\

Following the methods described in last section, several pieces of KCSO single crystals were measured by the micro-Hall magnetometer. Figure~\ref{KCSO}(a) shows the raw data of Hall resistance $R_{xy}$ as a function of field measured at 20~mK for a KCSO sample (sample size$\sim500\times500\times300$ $\mu$m $^{3}$; H$\parallel$ c-axis). The figure highlights the distinction between the channel carrying response from sample and the channel that was recognized as background, the former and the latter not only differ in line shape but also in amplitude. According to our discussion in last section, the intrinsic Hall response to the stray field $B_{\rm stray}$ of the sample is better presented after background subtraction. The so constructed $R_{xy}$ vs $\mu_0 H$ curve clearly reveals multiple plateaus reminiscent of magnetization plateaus expected for this compound, which to our knowledge have not been measured at such a low temperature thus far. A comparison of $R_{xy}(\mu_0H)$ measured at 1.6 K using the micro-Hall probe with the magnetization measured by MPMS at 2 K is presented in Fig.~\ref{KCSO}(b), the latter agreeing well with literature\cite{KCoSeO-Cava}. The fact that $R_{xy}(\mu_0H)$ can be scaled to $M(\mu_0H)$ verifies that the Hall resistance indeed reflects the magnetization of the sample. Furthermore, it shows that by data from direct magnetization measurements as anchor, a scaling factor, or a conversion coefficient, can be extracted, and then the total magnetization of a sample could at least semi-quantitatively be determined by the Hall magnetometer measurements.

To check the performance of the Hall device under intense magnetic field, a KCSO/device assembly, which had been measured in an $18$ T magnet, was further studied up to $30$ T in the all superconducting magnet at SECUF\cite{A3Station}. The results are shown in Fig.~\ref{KCSO}(c); both curves are background-subtracted and scaled by the corresponding $18$ T values to allow comparison. The overall agreement is apparent, though the actual Hall resistances differ by a factor of three as indicated in the legend of Fig.~\ref{KCSO}(c). Besides, for the $30$ T run, the first plateau is not as flat as that in the $18$ T run. We note that since these two measurements were conducted in two systems with different cryogenic refrigerators and superconducting magnets, the measurement condition other than temperature and field could be slightly different. For example, the probes used were both top-loading type with single axis rotator, the alignment of the sample/device assembly within the plane defined by magnetic field direction and the rotator axis could not be easily judged by the Hall sensor response itself, thus the full alignments during the two measurements could be off. Further study is needed to clarify to what extent the off-alignment in different systems could affect the response from the sample channel and the background channel and eventually the consistency of measurements. Despite this uncertainty, the differential d$R_{\rm xy}$/d$B$ shown in Fig.~\ref{KCSO}(d) captures all the magnetization features of KCSO when compared with previously reported magnetization result\cite{KCSO-arxiv}. \\

\begin{figure}
\includegraphics[width=0.9\columnwidth]{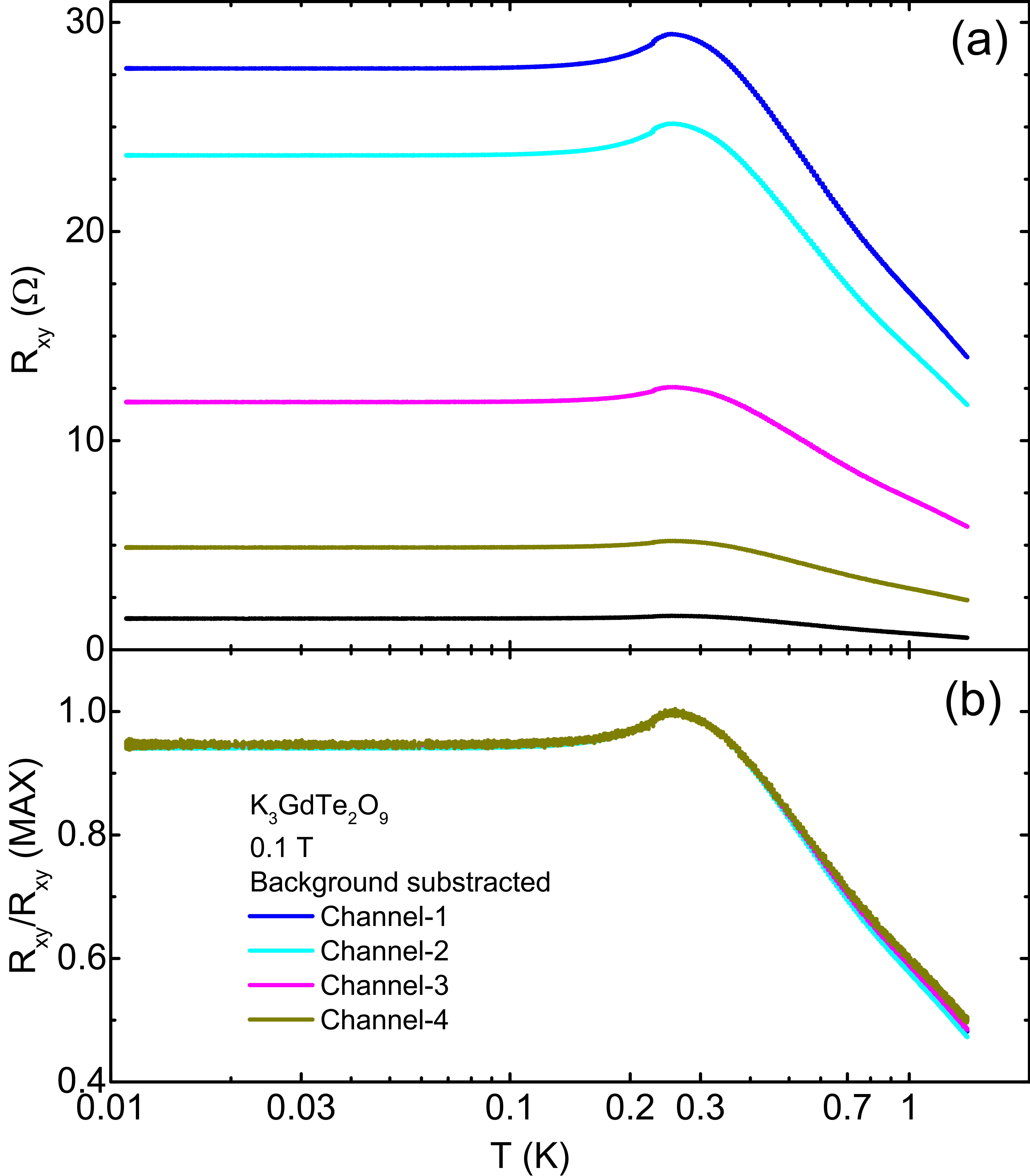}
\caption{\label{KGTO}
Magnetization of a  poly-crystalline K$_3$GdTe$_2$O$_9$ sample versus temperature at $0.1$ T,  measured in a dilution refrigerator. (a) is the raw Hall resistance of five Hall channels. The Hall crosses are further away from the sample with increasing channel number. (b) is the normalized data after subtracting the furthest channel as background.
}
\end{figure}

\noindent \textbf{Temperature-dependent magnetization.}
In addition to the field-dependent measurements, temperature-dependent magnetization has been also studied on a quantum magnet, K$_3$GdTe$_2$O$_9$, down to $\sim$ 15~mK by our micro-Hall device. This compound is a triangular-lattice frustrated antiferromagnet characterized by large lattice spacings and weak spin interactions. Previous specific-heat measurements revealed a phase transition at approximately 200 mK, yet this transition has not been confirmed by magnetization measurements, as such low temperatures are challenging to access in commercial magnetometers \cite{Ashtar2026}. Our measurements confirmed the antiferromagnetic transition via magnetization for the first time. As shown in Fig.~\ref{KGTO}(a), a broad peak is observed at 250 mK, followed by a sharp drop at 220 mK. As typically expected for frustrated magnets, the broad peak indicates the development of strong short-range spin correlations, while the sharp drop marks the onset of long-range antiferromagnetic order, as also seen in Ba$_2$La$_2$CoTe$_2$O$_{12}$ \cite{Kojima2018}. Furthermore, the Hall signals measured at four channels with different distances from the sample agree well after normalization (Fig.~\ref{KGTO}(b)). This not only confirms the effective background subtraction, but also demonstrates the reliability of the measurements from any channel, albeit with potentially different signal-to-noise ratios. \\

\begin{figure}
\includegraphics[width=0.99\columnwidth]{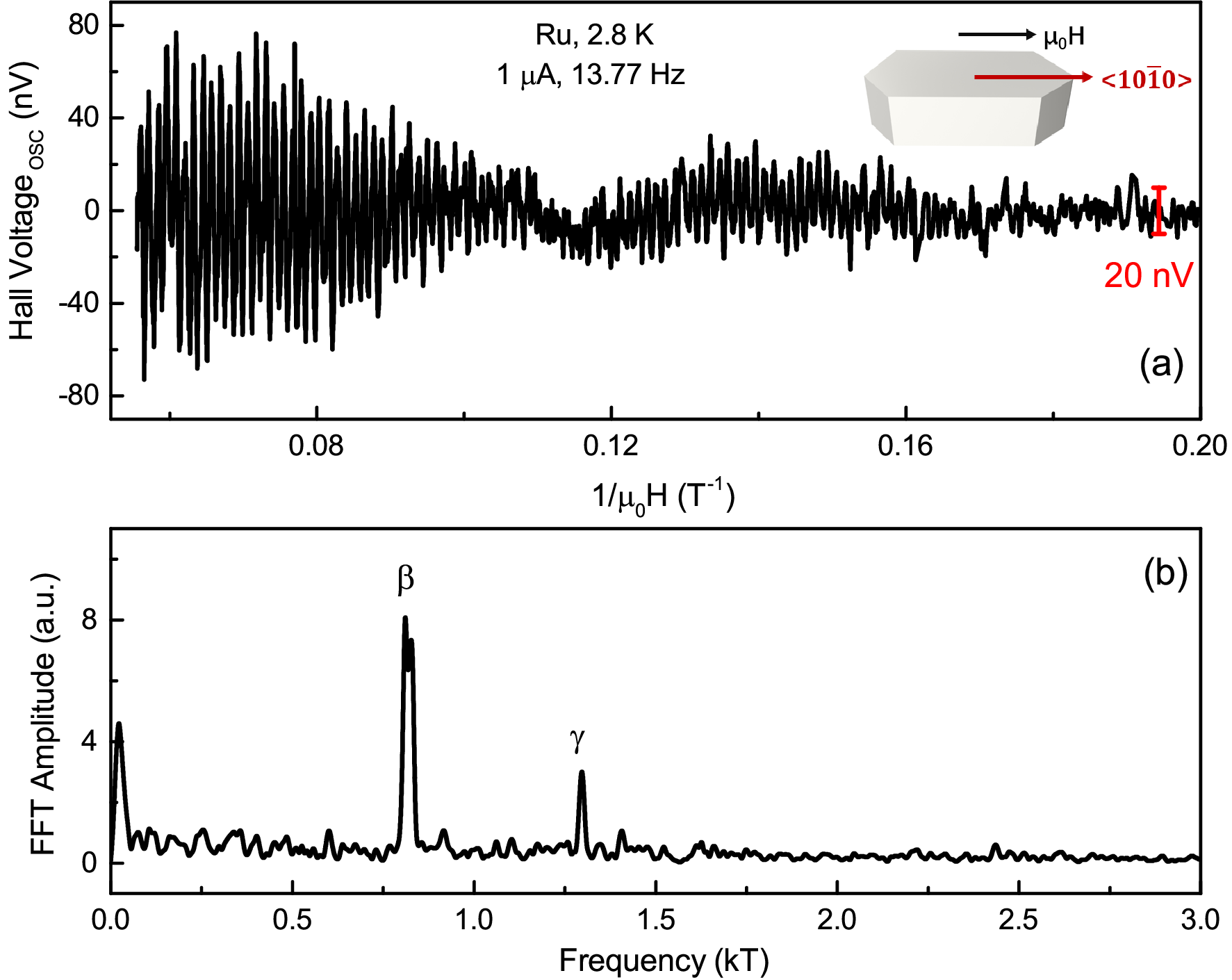}
\caption{\label{Ru-OSC}
The dHvA signal of a Ru single crystal detected by the Hall magnetometer. The lock-in amplifier was set with a time constant of 1~s and a slope of 18~dB/oct, and the field sweeping rate was 0.25~T/min. (a) The oscillatory component of the Hall voltage signal $V_{\rm osc}$ as a function of inverse magnetic field, after subtraction of a smooth background. The peak-to-peak voltage noise is below 20 nV. The crystal alignment with respect to the external magnetic field is shown as inset. (b) The corresponding FFT spectrum.
}
\end{figure}

\noindent \textbf{Quantum oscillations.}
We next employ the micro-Hall sensor to detect quantum oscillations in a material's magnetization as a function of applied magnetic field, i.e., the de Haas–van Alphen (dHvA) effect\cite{Shoenberg1984}, for which sufficiently low temperatures and high magnetic fields are usually prerequisites.
Conventionally, the ac-field modulation with pickup coil and the magnetic torque are widely adopted methods for dHvA measurements, and both have been implemented at dilution refrigerator temperatures and in high magnetic fields. By contrast, standard dc magnetization measurements by VSM are not well suited for detecting dHvA signals, owing to the difficulty of integrating a VSM under extreme conditions, its relatively slow response speed, and its often insufficient sensitivity. In the case of a Hall probe, it "transforms" the sample’s magnetization into a voltage signal measured in usual ac transport configuration, thus endowing rapid data acquisition during field sweeps. Given that the micro-Hall sensor responds to the total magnetization of a sample, whether the dHvA signal could be detected now depends on the sensor's sensitivity.

Elemental ruthenium (Ru), which crystallizes in the hexagonal close-packed (hcp) structure, was employed as a test sample. The measurements were performed with the magnetic field applied along the $<$10\=10$>$ crystallographic direction. Fig.~\ref{Ru-OSC}(a) presents the Hall signal given in nanovolts as a function of inverse field, both the reference channel Hall signal and a smooth background have been subtracted. The oscillations are clearly visible. The corresponding Fast Fourier transformation (FFT) analysis is shown in Fig.~\ref{Ru-OSC}(b); two distinct frequencies at about 0.8 kT and 1.3 kT were observed, which are in accordance with the $\beta$ and $\gamma$ closed-hole pockets in literature\cite{Coleridge1969}, respectively. The result confirms that the micro-Hall sensor can be used as a sensitive probe of dHvA effect. In the current setup, the sample's orientation with respect to the Hall sensor could not be varied \textit{in situ} during one cool down, so only temperature-dependent quantum oscillation study for a certain orientation could be readily performed. However, a two-axis rotator that adjusts the relative orientation of the sample to the external magnetic field while keeping the field parallel to the sensor plane could be conceived in future developments.\\

An estimate of field resolution of the Hall magnetometer can be made based on our available data. As indicated in Figure~\ref{Ru-OSC}(a), with an ac excitation current of 1~$\mu$A at 13.77 Hz, the lock-in measured peak-to-peak Hall voltage noise $\delta V_{\rm pp}$ $\approx$ 20 nV. Assuming Gaussian noise statistics, this corresponds to a root-mean-square (rms)
noise $\delta V_{\rm rms} = \delta V_{\rm pp}/6.6 \approx 3$ nV (a
$\pm 3.3 \sigma$ conversion enclosing 99.9\% of the fluctuations), and hence
$\delta R_{\rm rms} = 0.003~\Omega$. Given the micro-Hall sensor's field responsivity of 3.885~$\Omega$/mT, a field resolution $\approx$ 0.8 $\mu$T is obtained. This is comparable to the resolution achieved by ac-modulated quantum-well Hall sensors \cite{Kerlain2008} and conventional dc Hall magnetometers, which typically reach ~1 $\mu$T \cite{Haned2003}. Since we had MPMS data of the same piece of Ru crystal (not shown here), an estimate of magnetization resolution in the unite of emu is made as following. At $7$ T, the averaged magnetization by MPMS was 4E-4 emu, while by Hall magnetometer the background channel subtracted resistance was 14 $\Omega$, thus the conversion coefficient is 4E-4/14 = 2.9E-5 emu/$\Omega$, combining the resistance noise of $0.003~\Omega$, we have $\delta$M =  2.9E-5 $\times$ 3E-3 = 8.7 E-8 emu $\sim$ 1E-7 emu, which is at the reasonable level for a Hall magnetometer.\\

\section{Summary and Outlook}
To summarize, a micro-Hall sensor based on 2DEG in GaAs/AlGaAs heterostructure is designed and fabricated. It can be used as a magnetometer capable of working at temperatures down to $\sim$~15 mK and up by two orders to 20 K, as well as in steady field up to 30~T, partially fulfilling the demand for magnetization measurements under extreme conditions. Owing to its small footprint and convenient sample mounting procedure, the micro-Hall sensor can be easily integrated into existing cryogenic systems.

While our tests have demonstrated the robustness and sensitivity of the micro-Hall sensor magnetometer, there are still several aspects merit further exploration. Firstly, beyond the general-purpose micro-Hall chip, these parameters -- the field responsivity, the active area size, and spacings between Hall crosses -- could all be optimized according to the specific sample under study, to better accommodate high-sensitivity measurements at millikelvin temperature range and high magnetic field. Secondly, since the stray field of a magnetized sample decays cubically with distance, a multi-cross Hall sensor actually provides the spatial distribution of the stray field. Given that the relative position of the sample to the sensor is fixed, with the knowledge of sample geometry and demagnetization factor, it is possible to rebuild a sample's magnetization from Hall sensor readings, thus providing another method to obtain the magnetization value, other than using VSM data for scaling. Thirdly, it should be noted that all measurements in this work were performed in wet refrigerators, where samples were immersed in heat exchange gas or liquid, thus thermalization has not been a concern. In order to apply the micro-Hall sensor magnetometer to refrigerators with sample in vacuum, especially dry dilution refrigerators, the sample mounting scheme needs reconsideration, possibly by changing the sapphire piece to a silver bar that has direct contact with the fridge's cold finger. We expect that with further developments, the applicability of micro-Hall magnetometry can be significantly broadened, making it a routinely usable tool for magnetization studies under extreme environments.

\begin{acknowledgments}
We thank Xiaohui Song, Youguo Shi, Malik Ashtar, and Shiliang Li for providing high quality samples used in this work, and Changli Yang for fruitful discussions. This work was supported by the National Key Research and Development Program of China (Grant Nos. 2023YFA1607400, 2022YFA1602800, 2022YFA1402200, and 2021YFA0718700), the National Natural Science Foundation of China (Grant No. 12474147), and the Strategic Priority Research Program and Key Research Program of Frontier Sciences of the Chinese Academy of Sciences (Grant No. XDB1270102). This work is carried out at the Synergetic Extreme Condition User Facility (SECUF, https://cstr.cn/31123.02.SECUF).
\end{acknowledgments}

\bibliographystyle{aipnum4-1}   
\bibliography{Hall-M}

\end{document}